\documentclass[acmlarge]{acmart}
\AtBeginDocument{%
  }

\usepackage{longtable}
\setcopyright{acmlicensed}
\copyrightyear{2025}
\acmYear{2025}
\acmDOI{XXXXXXX.XXXXXXX}

\DeclareUnicodeCharacter{2500}{--}
\begin{document}

\title{Inclusivity of AI Speech in Healthcare: A Decade Look Back}

\author{Retno Larasati}
\email{retno.larasati@open.ac.uk}
\orcid{0000-0002-6412-2598}
\affiliation{%
  \institution{The Open University}
  \city{Milton Keynes}
  \country{United Kingdom}
}

\renewcommand{\shortauthors}{Larasati et al.}

\begin{abstract}
 The integration of AI speech recognition technologies into healthcare has the potential to revolutionize clinical workflows and patient-provider communication. However, this study reveals significant gaps in inclusivity, with datasets and research disproportionately favouring high-resource languages, standardized accents, and narrow demographic groups. These biases risk perpetuating healthcare disparities, as AI systems may misinterpret speech from marginalized groups. This paper highlights the urgent need for inclusive dataset design, bias mitigation research, and policy frameworks to ensure equitable access to AI speech technologies in healthcare.
\end{abstract}





\maketitle

\section{Introduction}
The healthcare sector has consistently sought technological innovations to enhance the quality of care and improve patient outcomes. One transformative area in this regard is the speech technology especially after the popularity of Artificial Intelligence (AI). AI speech technologies have transformed healthcare by enabling innovations in patient communication, diagnostics, and also treatment \cite{johnson2014systematic}. From voice-activated electronic health records (EHRs) to AI-powered diagnostic tools, these technologies have streamlined workflows and improved patient outcomes \cite{kwong2020empowering,izworski2004artificial,zhang2023intelligent}.

While technological advancements have made speech recognition systems more robust, challenges persist in achieving inclusivity. Biases in training data, limited linguistic diversity, and insufficient contextual adaptability hinder the equitable application of these technologies across various demographics \cite{latif2020speech, koenecke2020racial, martin2023bias}. Addressing such disparities is crucial, as healthcare systems increasingly adopt AI solutions to democratize access and improve care quality globally.

This paper examines the state of inclusivity in AI speech technologies over the past decade, focusing on three key areas: language inclusivity, demographic inclusivity, and accessibility for individuals with speech impairments based on the available speech datasets and research paper. By looking at existing research and dataset, this paper aims to highlight availability and identify gaps, to inform future directions in the domain and ensure equitable access to AI speech in healthcare. 




\section{Inclusivity in Speech Datasets for AI Speech Recognition}
In this section, we explore the inclusivity of AI speech recognition datasets, specifically open datasets published between 2015 and 2024. The search focused on datasets specifically designed for automatic speech recognition (ASR) and text-to-speech (TTS) synthesis, with an emphasis on their language coverage, accent diversity, demographic representation (gender, age, ethnicity), and inclusion of speech impairments. 

\begin{table}[h]
\resizebox{\textwidth}{!}{%
\begin{tabular}{|l|l|l|l|l|l|l|l|}
\hline
\textbf{Dataset Name} &
  \textbf{Language(s)} &
  \textbf{Accent(s)} &
  \textbf{\begin{tabular}[c]{@{}l@{}}Demographic \\ (Gender, Age, Ethnicity)\end{tabular}} &
  \textbf{\begin{tabular}[c]{@{}l@{}}Speech \\ Impairment\end{tabular}} &
  \textbf{Year} &
  \textbf{Category} &
  \textbf{Paper} \\ \hline
LibriSpeech     & English                                  & No           & No                      & No & 2015 & ASR & \cite{panayotov2015librispeech}    \\ \hline
GigaSpeech      & English                                  & No           & No                      & No & 2021 & ASR & \cite{chen2021gigaspeech}          \\ \hline
SPGISpeech      & English                                  & L1, L2       & No                      & No & 2021 & ASR & \cite{o2021spgispeech}             \\ \hline
Europarl-ASR    & English                                  & No           & Balance Male and Female & No & 2021 & ASR & \cite{garces2021europarl}          \\ \hline
EdAcc &
  English &
  9 English Accents &
  Balance Male and Female, Age, Ethnicity &
  No &
  2023 &
  ASR &
  \cite{sanabria2023edinburgh} \\ \hline
Speech Commands & English                                  & No           & No                      & No & 2018 & ASR & \cite{warden2018speech}            \\ \hline
VOICE           & English                                  & No           & Gender                  & No & 2018 & ASR & \cite{richey2018voices}            \\ \hline
CrowdSpeech     & English                                  & No           & Balance Male and Female & No & 2021 & ASR & \cite{pavlichenko2021crowdspeech}  \\ \hline
United-Syn-Med  & English                                  & No           & No                      & No & 2024 & ASR & \cite{banerjee2024high}            \\ \hline
EasyCom         & English                                  & No           & No                      & No & 2021 & ASR & \cite{donley2021easycom}           \\ \hline
AISHELL-1 &
  Mandarin Chinese &
  North, South,Others &
  Balance Male and Female &
  No &
  2017 &
  ASR &
  \cite{bu2017aishell} \\ \hline
AISHELL-2 &
  Mandarin Chinese &
  North, South, Other &
  Balance Male and Female, Age &
  No &
  2018 &
  ASR &
  \cite{du2018aishell} \\ \hline
AISHELL-3 &
  Mandarin Chinese &
  North, South, Other &
  Gender, 4 Age Groups &
  No &
  2020 &
  TTS &
  \cite{shi2020aishell} \\ \hline
AISHELL-4       & Mandarin Chinese                         & No           & Gender                  & No & 2021 & ASR & \cite{fu2021aishell}               \\ \hline
WenetSpeech     & Mandarin Chinese                         & No           & No                      & No & 2022 & ASR & \cite{zhang2022wenetspeech}        \\ \hline
MagicData-RAMC &
  Mandarin Chinese &
  North, South &
  Balance Male and Female &
  No &
  2022 &
  ASR &
  \cite{yang2022open} \\ \hline
CSRC            & Mandarin Chinese                         & No           & 3 Age Groups            & No & 2021 & ASR & \cite{yu2021slt}                   \\ \hline
THCHS-30        & Mandarin Chinese                         & No           & No                      & No & 2015 & ASR & \cite{wang2015thchs}               \\ \hline
LibriVoxDeEn    & German, English                          & No           & No                      & No & 2019 & ASR & \cite{beilharz2019librivoxdeen}    \\ \hline
MediaSpeech &
  Arabic, French, Turkish, and Spanish &
  No &
  No &
  No &
  2021 &
  ASR &
  \cite{kolobov2021mediaspeech} \\ \hline
ADIMA           & Arabic, and 9 other Indic languages      & No           & No                      & No & 2022 & ASR & \cite{gupta2022adima}              \\ \hline
ClovaCall       & Korean                                   & No           & No                      & No & 2020 & ASR & \cite{ha2020clovacall}             \\ \hline
VietMed         & Vietnamese                               & North, South & Gender                  & No & 2024 & ASR & \cite{le2024vietmed}               \\ \hline
BD-4SK-ASR      & Sorani Kurdish                           & No           & No                      & No & 2019 & ASR & \cite{qader2019kurdish}            \\ \hline
RTASC           & Romanian                                 & No           & Balance Male and Female & No & 2021 & ASR & \cite{puaics2021human}             \\ \hline
NPSC            & Norwegian Bokmål, Norwegian Nynorsk      & No           & Gender                  & No & 2022 & ASR & \cite{solberg2022norwegian}        \\ \hline
CoVoST          & Multilingual (29 Languages)              & 60 Accents   & No                      & No & 2020 & ASR & \cite{wang2020covost}              \\ \hline
CoVoST2 &
  Multilingual (16 Languages) &
  66 Accents &
  3 Genders, 8 Age Groups &
  No &
  2020 &
  ASR &
  \cite{wang2020covost2} \\ \hline
Europarl-ST     & Multilingual (9 European languages)      & No           & No                      & No & 2020 & ASR & \cite{iranzo2020europarl}          \\ \hline
MaSS            & Multilingual (8 European language)       & No           & No                      & No & 2019 & ASR & \cite{boito2019mass}               \\ \hline
OLR 2021        & Multilingual                             & No           & No                      & No & 2021 & ASR & \cite{wang2021olr}                 \\ \hline
VoxPopuli       & Multilingual (23 European languages)     & No           & No                      & No & 2021 & ASR & \cite{wang2021voxpopuli}           \\ \hline
NusaCrowd &
  Multilingual (19 Indonesian languages) &
  No &
  No &
  No &
  2022 &
  ASR &
  \cite{cahyawijaya2022nusacrowd} \\ \hline
MSNER           & Dutch, French, German and Spanish        & No           & No                      & No & 2024 & ASR & \cite{MSNER}                       \\ \hline
LibriTTS        & English                                  & No           & Balance Male and Female & No & 2019 & TTS & \cite{zen2019libritts}             \\ \hline
KazakhTTS       & Kazakh                                   & No           & Balance Male and Female & No & 2021 & TTS & \cite{mussakhojayeva2021kazakhtts} \\ \hline
RyanSpeech      & English                                  & No           & No                      & No & 2021 & TTS & \cite{zandie2021ryanspeech}        \\ \hline
SOMOS           & English                                  & No           & No                      & No & 2022 & TTS & \cite{maniati2022somos}            \\ \hline
SpeechInstruct  & Multilingual                             & No           & No                      & No & 2023 & TTS & \cite{zhang2023speechgpt}          \\ \hline
CVSS            & Multilingual (21 languages into English) & No           & No                      & No & 2022 & TTS & \cite{jia2022cvss}                 \\ \hline
EMOVIE          & Mandarin Chinese                         & No           & No                      & No & 2021 & TTS & \cite{cui2021emovie}               \\ \hline
\end{tabular}%
}
\caption{Speech Datasets}
\label{tab:data}
\end{table}

\subsection{Methods}
The search was conducted using PapersWithCode, a platform that links machine learning papers with their associated datasets and codes. The datasets included should be published between 2015 and 2024 to capture a decade-long evolution of inclusivity in AI speech datasets. Datasets were filtered for AI speech data, specifically ASR (Automatic Speech Recognition) and TTS (Text-to-Speech) based on their primary application. We assumed that ASR and TTS data are the most used in the healthcare context. Each dataset was then evaluated based on four inclusivity metrics:
\begin{itemize}
    \item Language Coverage: The languages represented in the dataset.
    \item Accent Diversity: The inclusion of regional or non-native accents.
    \item Demographic Representation: The consideration of gender, age, and ethnicity among speakers.
    \item Speech Impairment Inclusion: Whether the dataset includes samples from individuals with speech impairments.
\end{itemize}

\subsection{Results and Discussion}
The search yielded 38 datasets, which were analysed based on the four metrics mentioned above: language, accent, demographic, and speech impairment (See Table \ref{tab:data}). 


\textbf{Language Coverage}
Language diversity in speech recognition datasets is essential for ensuring fair and equitable AI applications across different populations. The majority of publicly available speech datasets are English-dominated (15 datasets), including LibriSpeech, GigaSpeech, SPGISpeech. While these datasets have advanced AI-powered automatic speech recognition (ASR), their monolingual nature limits inclusivity. However, the inclusion of non-English languages has grown in recent years, particularly for major Asian languages (See Table \ref{tab:data2}). For example, datasets such as AISHELL-1 through 4 provide extensive coverage of Mandarin, including regional accents. The NusaCrowd dataset also covers 19 Indonesian languages, significantly expanding representation in Southeast Asia. Some datasets are also multi-country-lingual datasets, such as, CoVoST2 (21 languages), VoxPopuli (22 European languages), and ADIMA (Arabic and 9 Indic languages), which have broadened the scope of AI speech recognition beyond English. These efforts demonstrate a growing recognition of the need for multilingual AI systems, particularly in regions with high linguistic diversity.

However, despite these advancements, significant gaps remain in the representation of smaller and marginalized languages. No native African languages are included in any dataset. While African accents of English (e.g., Nigerian, Ghanaian) are represented in EdAcc, this does not address the need for native language support. The exclusion of African languages limits the applicability of AI speech technologies in healthcare and could further increase the quality of care disparity \cite{turon2024infectious}. No dataset also includes Indigenous languages from the Americas, Australia, the Pacific Islands, or elsewhere, such as Najavo or Maori. This implied that indigenous communities, which often face significant healthcare disparities \cite{marrone2007understanding}, are excluded from the benefits of AI speech technologies. Compared to the European languages coverage (26 Languages), this imbalance reflects a broader trend in AI research and development to focus on languages that already have abundant linguistic resources, such as large text corpora, speech datasets, and computational tools, while neglecting languages that lack these resources.

\begin{table}[h]
\begin{tabular}{|l|l|l|l|l|l|l|l|}
\hline
\textbf{Region}      & \textbf{Languages Covered}                         \\ \hline
East Asia            & Mandarin Chinese, Japanese, Korean                 \\ \hline
South Asia           & 9 Indic languages, Tamil \\ \hline
Southeast Asia       & Indonesian (19 languages), Vietnamese, Thai        \\ \hline
Central Asia         & Kazakh                                             \\ \hline
Middle East          & Arabic, Persian, Turkish, Sorani Kurdish           \\ \hline
Africa               & None (except African English accents in EdAcc)     \\ \hline
Indigenous Languages & None                                               \\ \hline
Europe               & 26 European languages                              \\ \hline
\end{tabular}
\caption{Non-English Only Datasets}
\label{tab:data2}
\end{table}

\textbf{Accent Inclusivity.} Accent diversity is critical for ensuring accurate speech recognition across regional and non-native speakers. While many speech datasets do not specify regional accent representation, some datasets explicitly capture accent diversity. For instance, EdAcc covers South African, Ghanaian, Irish, Scottish, US, Southern British, Indian, Jamaican, and Nigerian English accents. Similarly, AISHELL-1 and AISHELL-2 represent different Mandarin accents. These efforts improve recognition fairness, but most English ASR datasets ignore non-standard or non-native accents, leading to potential bias in AI models.

\textbf{Demographic Representation.} It is critical for AI models to perform equitably across different groups, such as different genders, ages, and ethnicities. Gender balance in speech datasets is critical for ensuring that AI models perform equally well for all speakers, regardless of their gender. However, many widely used datasets, such as LibriSpeech and GigaSpeech, do not specify the gender distribution of their speakers. This lack of transparency makes it difficult to assess whether these datasets are balanced or biased toward a particular gender. Some datasets have made efforts to address gender inclusivity. For example, Europarl-ASR and AISHELL-1 include balanced male and female speakers, ensuring that their models are trained on a representative sample of voices. Similarly, EdAcc balances gender representation and includes additional demographic factors such as age, race/ethnicity, and education level, making it one of the most inclusive datasets in terms of demographic representation. However, the representation of non-binary or gender-diverse individuals remains almost entirely absent. Only one dataset, CoVoST2 (2020), explicitly includes more than two genders, covering three gender categories. This is a significant step forward, but it remains an exception rather than the norm. The lack of gender diversity in most datasets limits the ability of AI systems to serve gender-diverse populations, particularly in healthcare settings where inclusive communication is critical.

Age diversity in speech datasets is another critical factor, as speech patterns can vary significantly across different age groups. Older adults, in particular, are often underrepresented in AI speech datasets, despite being one of the populations that could benefit most from speech technology \cite{klaassen2024review}. Datasets like AISHELL-3 and CoVoST2 have made efforts to include multiple age groups, with CoVoST2 covering eight age categories. However, these datasets are exceptions. Another example is CSRC, which explicitly includes speech samples from children. This is a significant step forward, as children have distinct speech patterns that differ from adults. Most datasets, such as LibriSpeech, do not specify the age distribution of their speakers, making it difficult to assess whether they include older adults or children.

Lastly, the ethnic and racial demography. Most of the datasets do not specify the ethnic or racial composition of their speakers, leading to potential biases in AI models. For example, datasets trained primarily on Western or European voices may struggle to accurately recognize speech from individuals with non-Western accents or dialects. Some datasets have made efforts to address ethnic and racial diversity. EdAcc included race and ethnicity in the data, making it one of the few datasets to explicitly address this issue. Other datasets (e.g. CoVoST), which included different accents from diverse linguistic and cultural backgrounds, might have different race and ethnicity in the data though it does not provide detailed metadata on ethnicity.

\textbf{Speech Impairment Inclusion.} One of the most significant gaps in AI speech recognition datasets is the lack of representation for individuals with speech impairments. Despite the increasing need for accessible AI speech tools for people with disabilities, none of the datasets we found explicitly include impaired speech samples. This exclusion creates major barriers in AI accessibility, as speech recognition systems often perform poorly for individuals with speech impairments. The absence of datasets focusing on dysarthria, aphasia, or other speech disorders underscores the urgent need for dedicated speech impairment datasets to train more inclusive AI models.

\subsection{Limitations}
While the dataset search provides valuable insights into inclusivity trends, it has some limitations. The analysis is limited to datasets available on PapersWithCode, which may exclude proprietary or less known datasets, or datasets without open access papers. Some datasets lack detailed metadata on accent, demographic, or speech impairment inclusion, making it challenging to assess their inclusivity fully. 


\section{Research on Inclusive Speech AI in Healthcare}
To analyze research trends in AI speech recognition for healthcare, we conducted a systematic search using OpenAlex, a comprehensive scholarly database. The search strategy was designed to capture two broad categories of research: (1) general speech technology in healthcare, serving as a control to establish the overall volume of research in the field, and (2) inclusive AI speech recognition, focusing on studies addressing bias and demographic inclusivity. All searches were restricted to publications between 2015 and 2024 and utilized three key parameters: full-text terms, title/abstract keywords, and domain (Health Sciences). 

\subsection{Methods}
For the control search, we identified papers broadly related to speech technology in healthcare using the full-text term "AI speech technology healthcare" within domain "Health Sciences." This search aimed to quantify the total number of publications in the field, providing a baseline for comparing the volume of inclusivity-focused research.

To investigate inclusive AI speech recognition, we conducted targeted searches combining full-text terms, title/abstract keywords, and the domain filter (Health Sciences). For example, studies addressing bias in AI speech recognition were identified using the full-text term "inclusive AI speech healthcare bias," while demographic-specific research was captured by pairing this term with title/abstract keywords from different demographic representations. For gender representation, the search keywords are: gender, women, men, transgender, non-binary, gender-diverse. For age inclusiivity, the search keywords are: age, elderly, geriatric, paediatric. For Speech Impairment, the search keywords are: speech impairment, speech disorders, dysarthria, aphasia. For racial bias, the search keywords are: ethnic, racial, cross-cultural, indigenous.

The domain parameter (Health Sciences) ensured that results were confined to healthcare applications, while full-text and title/abstract parameters balanced breadth and precision. For instance, the search for gender inclusivity combined the full-text term "inclusive AI speech healthcare bias" with the title/abstract keyword "gender" and the domain filter. This approach minimized off-topic results while capturing studies that explicitly addressed demographic biases. By comparing the volume of general AI speech technology papers (control search) to inclusivity-focused research, we assessed how prioritization of fairness and diversity has evolved over the past decade. The results are shown in Table \ref{tab:OA}. 

\begin{table}[]
\resizebox{\textwidth}{!}{%
\begin{tabular}{|l|l|llllll|}
\hline
search criteria &
  full text: AI speech technology healthcare &
  \multicolumn{6}{c|}{full text: inclusive AI speech technology healthcare bias} \\ \cline{2-8} 
 &
  \multicolumn{1}{c|}{domain: Health Sciences} &
  \multicolumn{1}{l|}{} &
  \multicolumn{5}{c|}{domain: Health Sciences} \\ \cline{1-2} \cline{4-8} 
year &
   &
  \multicolumn{1}{l|}{} &
  \multicolumn{1}{l|}{} &
  \multicolumn{1}{l|}{title:gender} &
  \multicolumn{1}{l|}{title:age} &
  \multicolumn{1}{l|}{title:speech impairment} &
  title:ethnic \\ \hline
2024 &
  676 &
  \multicolumn{1}{l|}{479} &
  \multicolumn{1}{l|}{97} &
  \multicolumn{1}{l|}{0} &
  \multicolumn{1}{l|}{12} &
  \multicolumn{1}{l|}{7} &
  0 \\ \hline
2023 &
  979 &
  \multicolumn{1}{l|}{676} &
  \multicolumn{1}{l|}{120} &
  \multicolumn{1}{l|}{2} &
  \multicolumn{1}{l|}{14} &
  \multicolumn{1}{l|}{5} &
  2 \\ \hline
2022 &
  437 &
  \multicolumn{1}{l|}{210} &
  \multicolumn{1}{l|}{28} &
  \multicolumn{1}{l|}{0} &
  \multicolumn{1}{l|}{2} &
  \multicolumn{1}{l|}{0} &
  0 \\ \hline
2021 &
  374 &
  \multicolumn{1}{l|}{159} &
  \multicolumn{1}{l|}{26} &
  \multicolumn{1}{l|}{1} &
  \multicolumn{1}{l|}{3} &
  \multicolumn{1}{l|}{0} &
  1 \\ \hline
2020 &
  216 &
  \multicolumn{1}{l|}{86} &
  \multicolumn{1}{l|}{9} &
  \multicolumn{1}{l|}{0} &
  \multicolumn{1}{l|}{1} &
  \multicolumn{1}{l|}{0} &
  0 \\ \hline
2019 &
  159 &
  \multicolumn{1}{l|}{84} &
  \multicolumn{1}{l|}{13} &
  \multicolumn{1}{l|}{0} &
  \multicolumn{1}{l|}{2} &
  \multicolumn{1}{l|}{1} &
  0 \\ \hline
2018 &
  78 &
  \multicolumn{1}{l|}{44} &
  \multicolumn{1}{l|}{4} &
  \multicolumn{1}{l|}{0} &
  \multicolumn{1}{l|}{0} &
  \multicolumn{1}{l|}{0} &
  0 \\ \hline
2017 &
  32 &
  \multicolumn{1}{l|}{11} &
  \multicolumn{1}{l|}{0} &
  \multicolumn{1}{l|}{0} &
  \multicolumn{1}{l|}{0} &
  \multicolumn{1}{l|}{0} &
  0 \\ \hline
2016 &
  25 &
  \multicolumn{1}{l|}{7} &
  \multicolumn{1}{l|}{3} &
  \multicolumn{1}{l|}{0} &
  \multicolumn{1}{l|}{0} &
  \multicolumn{1}{l|}{0} &
  0 \\ \hline
2015 &
  21 &
  \multicolumn{1}{l|}{1} &
  \multicolumn{1}{l|}{1} &
  \multicolumn{1}{l|}{0} &
  \multicolumn{1}{l|}{0} &
  \multicolumn{1}{l|}{1} &
  0 \\ \hline
\end{tabular}%
}
\caption{Numbers of Paper from 2015-2024 via OpenAlex. Three search filters were used, such as full text, domain, and title/abstract.}
\label{tab:OA}
\end{table}

\subsection{Results and Discussion}
The control search, which identified papers broadly related to speech technology in healthcare using the full-text parameter: "AI speech technology healthcare," demonstrates a significant increase in research activity over the past decade. In 2015, only 21 papers were published in this area, but by 2024, the number had surged to 676. This nearly 32-fold growth reflects the rising importance of AI speech technologies in healthcare, driven by advancements in machine learning and the increasing adoption of voice-based tools in clinical settings, with the peak in 2023 (979 papers). 

Research explicitly addressing inclusivity and bias in AI speech recognition, identified using the full-text term "inclusive AI speech technology healthcare bias," has also grown significantly, though it remains a small fraction of the overall field. In 2015, only one paper focused on inclusivity, but by 2024, this number had risen to 479. Despite this growth, inclusivity-focused research accounted for just 10\% of all speech technology papers in 2024, highlighting the need for greater prioritization of fairness and diversity in AI development. Following the trend from the previous search, the peak was shown in 2023, with 676 papers, and then decline to 479 papers in 2024. This suggests that inclusivity remains a niche area within the field, requiring sustained effort to address persistent gaps. Especially when we narrow down the search further, with Health Sciences domain. The peak was in 2023 with only 120 papers, and down to 97 papers in 2024. 

\textbf{Gender Representation.} Research on gender inclusivity in AI speech recognition, identified using gender related keywords in the title/abstract search remains strikingly limited. In 2024, no papers explicitly addressed gender bias, and only two papers were published in 2023. This lack of research is concerning, given documented disparities in AI recognition accuracy for male versus female speakers and the growing recognition of gender diversity in healthcare. The absence of gender-focused research in 2024 underscores the need for greater attention to this issue, particularly as AI speech technologies become more widely used in clinical settings where equitable communication is critical.

\textbf{Age Inclusivity.} Research on age-related biases in AI speech recognition, identified using the title/abstract keywords related to different age group has seen modest growth but remains limited. In 2024, 12 papers addressed age inclusivity, up from just one in 2021. However, this represents a small fraction of the overall research on AI speech technologies. Only 26 papers in 2023 and 2024 explicitly addressed elderly-specific AI speech research, despite the growing need for speech technologies that cater to older adults, particularly in healthcare settings where age-related conditions such as hearing loss or speech impairments are common. Unfortunately, research on pediatric speech recognition remains scarce.

\textbf{Speech Impairment and Disability.} Research on speech impairments and disabilities, is also critically lacking. In 2024, only seven papers addressed speech impairments, and five papers in 2023. This gap is particularly concerning, as individuals with speech impairments or disabilities are among the populations that could benefit most from AI speech technologies. 

\textbf{Ethnic and Racial Bias Inclusivity.} Research on ethnic and racial biases in AI speech recognition remains scarce. In 2023, only two papers were retrieved, and the total number of papers over the past decade is minimal. This lack of research can be problematic, as AI models trained on Western-centric datasets often struggle to recognize speech from individuals with non-Western accents or dialects, exacerbating healthcare disparities.

\subsection{Limitations}
While the OpenAlex search provides insights
into research trends, it has some limitations. The analysis is limited to publications indexed in OpenAlex, which may exclude relevant studies from other databases. Moreover, the reliance on specific keywords may miss studies that address inclusivity and bias without explicitly using the searched terms. For example, the domain filter might excluded papers in healthcare area that are not mainly categorised as "Health Sciences".


\section{Conclusion}
The integration of AI speech recognition technologies into healthcare holds immense promise for improving clinical workflows, patient-provider communication, and accessibility. However, this study’s analysis of research trends (via OpenAlex) and speech datasets (via PapersWithCode) reveals persistent gaps in inclusivity that threaten equitable access to these advancements. Over the past decade, while the volume of research on AI speech technologies in healthcare has surged—growing from 21 papers in 2015 to 676 in 2024—only a small fraction explicitly addresses inclusivity or bias. Similarly, speech datasets remain skewed toward high-resource languages, standardized accents, and narrow demographic groups, failing to represent the diversity of global populations.

These gaps shown negative implications for healthcare equity. AI systems trained on biased datasets risk misinterpreting speech from marginalized groups—older adults, non-native speakers, or individuals with speech disorders, could lead to errors in clinical documentation, misdiagnoses, or inadequate patient engagement. The lack of representation in research further entrenches these disparities. This results also highlights possible areas of progress. Datasets like EdAcc (2023) and CoVoST2 (2020) demonstrate that intentional design can improve accent and demographic diversity, while research on inclusivity, though limited, has grown steadily since 2020. 

To ensure AI speech technologies serve all populations equitably, future researcher should expand datasets for underrepresented languages such as African or Indigenous languages, partnering with local communities to ensure ethical data collection. It is also crucial for future researcher to consider gender, age, and ethnic diversity during the dataset design, with transparent metadata to audit representation. Future studies could also develop datasets and models that capture diverse speech patterns, including dysarthria, aphasia, and other speech disabilities. By centering inclusivity in dataset design, research development, and technological advancement, we can build AI tools that truly democratize access to healthcare, empowering every patient and provider, regardless of language, accent, age, or ability.


\bibliographystyle{ACM-Reference-Format}
\bibliography{sample-base}

\end{document}